\documentclass[aps,showpacs,floatfix,a4paper]{revtex4}
\usepackage{graphicx}
\usepackage{amssymb,amsmath}
\usepackage{textcomp}

\begin{document}

\title{ Comment on 
``Electron transport through correlated molecules computed using the time-independent Wigner function: 
Two critical tests"}

\author{J.C. Greer$^1$}
\email{jim.greer@tyndall.ie}
\author{P. Delaney$^2$}
\email{p.delaney@qub.ac.uk}
\author{ G. Fagas$^1$}
\email{georgios.fagas@tyndall.ie}
\affiliation{$^1$Tyndall National Institute, 
University College Cork, Lee Maltings, Prospect Row, Cork, Ireland \\
$^2$School of Mathematics and Physics, Queen's University Belfast, Northern Ireland BT7 1NN}
\date{\today}

\begin{abstract}
The many electron correlated scattering (MECS) approach to
quantum electronic transport was investigated in the linear response regime
[I. B{\^a}ldea and H. K{\"o}ppel, Phys. Rev. B. {\bf 78}, 115315 (2008)]. The authors
suggest, based on numerical calculations, that the manner in which the method 
imposes boundary conditions is unable
to reproduce the well-known phenomena of conductance quantization. We introduce an 
analytical model and demonstrate that conductance quantization
is correctly obtained using open system boundary conditions within the MECS
approach. 
\end{abstract}

\pacs{72.10Bg,72.90.+y,73.23Ad}
\maketitle

{\section{Introduction}
\label{I}}

Electron transport calculations in the linear response regime for a 
method referred to as the stationary Wigner function (SWF) method have been reported 
[I. B{\^a}ldea and H. K{\"o}ppel, Phys. Rev. B. {\bf 78}, 115315 (2008)], as a shorthand
we will refer to their presentation as BK.  In fact though, their paper is a comment on 
the validity of a method introduced by Delaney and Greer~\cite{DeG04a} for treating open 
system boundary conditions for correlated many-electron problems, or many-electron correlated 
scattering (MECS)~\cite{DeG04a,ASS06,MGF09}. For reasons that will hopefully become clear, 
we refer to the method under discussion as MECS~\cite{SWF}. BK express several criticisms of the MECS 
method, but their primary objection and conclusion is 

{\it ``... the manner in which it" } (open system boundary conditions through the
Wigner function) {\it ``was imposed in"} [P. Delaney and J.C. Greer, Phys. Rev. Lett. {\bf 93},
036805 (2004)] {\it ``turns out to be inappropriate. It misses the fact that, in accord with 
our physical understanding, the current flow is due to an asymmetric injection of 
electrons from reservoirs into the device, and that injected electrons are very well 
described by Fermi distributions with different chemical potentials. Moreover, as results
from the analysis at the end of Sec. VI"} (of the BK paper), {\it ``unfortunately there is 
no simple remedy of the SWF method; the modification of the boundary conditions in the spirit of"} 
Delaney and Greer {\it ``such as to account for a nonvanishing chemical potential shift does not 
yield the desired improvement."}

The conclusion in BK is reached after numerical calculations on a test system designed to model 
independent electron and correlated electron transport across a quantum dot. As we will demonstate, 
their conclusion incorrectly presumes an asymmetric injection of momentum is necessary to describe
reservoirs or electrodes in quasi-equilibrium, and likewise incorrectly assumes there is no chemical 
potential difference when applying the open system boundary conditions with the MECS procedure. 
We will show that the Wigner boundary conditions as expressed for MECS calculations, or equally in 
previous works~\cite{Fre90}, {\it is} consistent  with the application of a ``{\it non-vanishing 
chemical potential shift}". The authors in their paper appear to be confusing the application of 
open boundary conditions when using either energy or momentum distributions, and we will clarify that
a relative shift in energy to the reservoir energy distributions {\it does not} imply a shift for 
the corresponding momentum distributions emitted into a scattering region for simple models of 
electrode behaviour. 

In sect.~\ref{II}, we give a brief overview of the MECS method. This is followed in 
sect.~\ref{III} by introduction of a resolution to the apparent conundrum expressed by 
BK: we analyse momentum distributions for electron reservoirs or electrodes represented by 
Fermi-Dirac  distributions and observe that the momenta distributions {\it do not} change with 
applied voltage bias, or equivalently with a chemical potential imbalance applied between electrodes. 
Asymmetric injection implied by BK's conclusion in relation to the momentum distributions 
from the electrodes is not consistent with a MECS or a Landauer description 
of electron transport. In sect.~\ref{IV}, a calculation of conductance quantization for a system with the
electrodes represented by parabolic energy bands in one spatial dimension is given using the MECS 
construction. To clearly identify how the boundary conditions can be treated in this case, electron 
interactions are neglected and the method reduces to a many-electron scattering problem for non-interacting 
electrons. This approximation has the advantage of highlighting the simplicity of the MECS formulation as
well as clearly demonstrating its consistency with standard formulations of electron scattering, 
while avoiding issues related to specific numerical approximations or questions related
to specific electronic structure implementations. The boundary conditions as formulated in 
MECS applied to the model reproduces the well-known result of conductance quantization.

{\section{A brief introduction to the MECS approach}\label{II}}

The proposal behind MECS as introduced in 
ref.~\cite{DeG04a} is to constrain a many-electron wavefunction on a scattering region, or 
specifically, the many-electron density matrix (DM) 
\begin{equation}
\rho_N = |\Psi_N><\Psi_N |,
\end{equation}
in a manner satisfying open system boundary equations~\cite{Fre90}. As open system boundary conditions
are commonly expressed in the language of single particle theories,
it is useful to consider the Wigner transform of the one-electron reduced density matrix (RDM) associated 
with the many-electron DM
\begin{eqnarray}
\rho_N &\rightarrow& \rho_1, \,\,\,\, {\rm with}\,\,\,\, {\rm Tr} \rho_1 = N \nonumber \\
f_W(q,p) &=& \int_{-\infty}^{+\infty} dr \, \exp (-ipr/\hbar)\rho_1(q-r/2;q+r/2),
\end{eqnarray}
where here the transform is written for one spatial dimension $r$ and $\{p,q\}$ are 
Wigner phase space variables.  The MECS approach is 
the recognition that open system boundary conditions can be applied to correlated systems 
through the one-body RDM through use of the Wigner transform, allowing appropriate conditions to be enforced
at the boundaries of a scattering region. In practice, the Wigner distribution function is 
used to constrain the momenta flow out of the electron reservoirs and into the scattering region.
The momentum expection value can be written as
\begin{equation}
\label{Waverage}
<p> = \frac{1}{2\pi\hbar}\int_{-\infty}^{+\infty} dp\, dq \,\, p \, f_W(q,p).
\end{equation}
Eq.~\ref{Waverage} highlights the analogy of the Wigner quantum phase space distribution to a classical
probability distribution function. However, unlike a classical probability distribution and
as is well known~\cite{Fre90,CaZ83,HOS84,Lee95}, the Wigner function is not everywhere positive as a
consequence of the Heisenberg momentum-position uncertainty principle. However, in regions where $f_W$ 
behaves approximately classically, 
the Wigner phase representation allows us to assign meaning to phrases such as 
``electrons in the left or right reservoir", 
``momentum of an electron emitted from a reservoir", or ``a reservoir is locally in equilibrium". 
Within this context, the net momentum flow out of the left electrode is approximated as
\begin{equation}
p_l = \frac{1}{2\pi\hbar}\int_0^{+\infty} dp \,\, p \,\, f_W(q_l,p),
\end{equation}
and similarly for the right electrode
\begin{equation}
p_r = \frac{1}{2\pi\hbar}\int_{-\infty}^0 dp \, \, p \,\, f_W(q_r,p)
\end{equation}
where $q_l$ and $q_r$ are appropriately chosen. Clearly this approximation is dependent upon how well
$f_W$ describes a classical probability distribution function. For metal electrodes, where electrons
can be well approximated by the free electron model, this assumption is usually justified. It is also
noteworthy that the Wigner reduced one-particle function, as a function of energy defined in the
Wigner phase space, tends rapidly toward the Fermi-Dirac distribution 
with increasing number of particles in a confining potential~\cite{CBJ07}.

In our calculations to date, three-dimensional electrodes are considered~\cite{DeG04a,MGF09,FDG06,FaG07}.
To simplify the analysis, $f_W$ is integrated over the in-plane coordinates within a cross section of 
the electrodes. The net momentum flow out of both electrodes is constrained to this equilibrium ($V=0$) 
value~\cite{DeG04a,DeG04b}. To consider this procedure further, we examine models of
electrode behaviour in quantum transport theories.

{\section{Fermi-Dirac reservoir energy and momentum distributions under applied voltage at zero temperature}
\label{III}}

In fig.~\ref{fig1}a), a pictorial representation of open system boundary conditions is shown
for electrodes described by two parabolic energy bands (free-electrons
with effective mass $m^*$ in one dimension and for a Fermi-Dirac distribution at temperature $T=0$).
The energy levels $\epsilon_k=\frac{(\hbar k)^2}{2m^*}$ in the left and 
right electrodes are filled to the Fermi energy  $\epsilon_F$; similarly
momentum states are filled to the Fermi momentum $k_F$. In fig.~\ref{fig1}b), a potential energy difference
is introduced between the left and right electrodes shifting the bottom of the energy bands with
respect to each other by an amount denoted $eV$. This results in a shift to the energies in the right
electrode by $\frac{(\hbar k)^2}{2m^*} \rightarrow \frac{(\hbar k)^2}{2m^*} +eV$ (a symmetric
split in the voltage between electrodes does not alter our discussion). The 
shift of the right electrode energy states describes the chemical potential imbalance between the 
reservoirs  and will necessarily be accompanied by a voltage drop, or equivalently an
electric field across the scattering region. In fig.~\ref{fig1}c), the momentum distributions corresponding 
to Fermi-Dirac energy distributions with and without applied voltage in fig.~\ref{fig1}a)
and fig.~\ref{fig1}b), respectively, are shown. The {\it same} momentum 
distributions are obtained with or without application of a voltage difference between the left and 
right electrodes. It also follows that similar considerations hold for the case of non-zero temperature.
It is important to highlight at this point that the MECS proposal~\cite{DeG04a} 
for treating quantum electronic transport is completely compatible with this picture of
boundary conditions on the electrode regions. 

When working with the one-electron RDM
obtained from a correlated $N$-electron density matrix, it is not possible to unambiguously 
define single electron energies. It is therefore of advantage to constrain the total incoming
momentum to model the action of the electrodes while determining the correlated 
many-electron density matrix on the scattering region. As the incoming momentum
distributions are the same for electrodes in equilibrium or in local equilibrium,
constraining the net momentum flow from the electrodes is not
sufficient to drive the electrodes away from equilibrium with respect to each other.
It is standard practice in many electron transport methods, including non-equilibrium Green's 
function (NEGF) techniques, to introduce an external electric field 
to model the action of the electrodes on the scattering 
region. To understand the role of the external field in transport 
calculations, it is useful to consider the sketch of two metallic electrodes as presented in fig.~\ref{fig2}.
A simple band or independent particle model is a remarkably good approximation to the electronic 
structure of metals allowing our discussion of the parabolic bands to be extended directly to consideration 
of realistic models of metal electrodes.  Within fig.~\ref{fig2}, the fact that the two electrodes 
are not in equilibrium with respect to one another is denoted by the different left $\mu_L$ and right $\mu_R$
chemical potentials. The chemical potential imbalance introduces a difference in the charge density 
between the left and right electrodes. However, electrostatic screening is efficient in metals and for 
the quasi-equilibrium regions of the electrodes, the electric field is zero within the electrodes
or equivalently, the voltage is constant within a short distance into the metal electrodes. Thus all of the voltage
drop is across the scattering region plus the screening length into the electrodes. Typical screening lengths 
in metals are of the order of 0.1 nm. Hence any charge imbalance in the electrodes resides at 
the surface of the metal and the opposite polarity of the surface induced charges between the electrodes 
gives rise to an electric field across the region situated between the electrodes;  
a situation depicted in fig.~\ref{fig2} as field lines between the electrode surfaces. 
In most transport calculations, charges in the electrode and scattering
regions are solved for self-consistently allowing a molecular tunnel junction to polarize in
response to this external electric field. In MECS, charges re-arrange due to 
minimization of the energy with respect to the many-electron wavefunction subject to the open system
boundary conditions and the external electric field. The voltage can then 
be extracted as the combined field arising from the applied field and polarized charge distribution 
in the scattering region; see for example~\cite{BMO02}. 

The model of electrode behaviour we are describing is consistent with
other formulations of quantum electronic transport. In this regard, it is worthwhile to mention
the work of McLennan, Lee and Datta~\cite{MLD91} and in particular their fig. 3. As pointed
out by the authors, a change in the chemical potential in the electrodes is accompanied by a shift
in the electrostatic voltage. In a metal, the case considered for MECS calculations to date,
the chemical potential and electrostatic voltage changes are nearly identical and the effect of
the applied voltage can be accounted for as a shift in the electrode bands with respect to each
other. In the linear
response regime, the shift due to the electrostatic voltage is sometimes neglected~\cite{MLD91}
but formally should be included as indicated schematically in our fig.~\ref{fig1} and in
McLennan {\it et al}'s fig. 3. 
 
The point which we would like to emphasize is that the presence of voltage drop on the scattering 
regions ensures that the single electron energies in the electrodes are shifted by the applied
voltage, but does not alter the momentum distributions 
emitted from the electrodes. The conclusion in BK that an asymmetric injection of electrons is 
needed to obtain a current is incorrect, if injection refers to incoming electron momentum distributions, 
i.e. `current  injected' from the electrodes. What is required is asymmetric scattering for injected 
electrons, and this is provided for by the asymmetric electric field profile or equivalently, the chemical
potential imbalance generated across a molecular tunnel junction.

{\section{Application of MECS to a single particle model}
\label{IV}}

\subsection{Boundary conditions}

MECS was originally formulated for interacting many-electron systems with a Hamiltonian 
operator given as
\begin{equation}
H_N = T + V_1 + V_2
\end{equation}
where $T$ is the sum of $N$ one-electron kinetic energy operators, $V_1$ is the sum of the
external one-electron potentials and $V_2$ is the sum of electron-electron interactions on 
the scattering region~\cite{DeG04a}. As mentioned, with the MECS approach an external electric 
field can be applied to model the chemical potential imbalance between the reservoirs and 
this term may be included into $V_1$. To arrive at a single particle model to be used in our
analysis, the electron-electron interactions are `switched' off and, for simplicity, all other 
external potentials other than the chemical potential imbalance between 
the electrodes are also `switched' off. As electron-electron interactions are not treated, the 
resulting many-electron system consists of $N$ non-interacting electrons each described by a single 
electron Hamiltonian operator. Of course, there is no advantage to apply the MECS method to a 
non-interacting electron model, but we demonstrate in this case that the MECS method 
is consistent with the usual formulation of quantum mechanical scattering.

We consider a scattering region of length $[-L/2,L/2]$ 
in the absence of a potential and write 
plane wave eigenfunctions on the scattering region
\begin{eqnarray}
\psi_n(x) &=& \frac{1}{\sqrt{L}} \exp(ik_nx) \nonumber \\
k_n &=& 2\pi n/L \,\,\,\,\,\,;\,\,n=\pm 1,\pm 2, \pm 3, \ldots
\end{eqnarray}
Left going and right going states are filled to a Fermi momentum $k_F=n_F\Delta k$. In the absence 
of the application of a voltage, the current is given simply as
\begin{equation}
I\left[ V=0 \right]=-\frac{e\hbar}{m^*L}\left( \sum_{n_l=1}^{n_F} k_{n_l}-\sum_{n_r=1}^{n_F} k_{n_r}\right)=0,
\end{equation} 
where $n_l$ and $n_r$ denote states associated with the left and right electrodes, respectively. 
In one dimension, the electron current and current density are equivalent, and the factor $\frac{1}{L}$ 
indicates our choice of normalization. The model consists of left and right 
propagating plane wave states with the net current summing to zero.  
In this representation, the density matrix is diagonal with 
$n_{n,n^\prime}= \delta_{nn^\prime}$ for both $n,n^\prime\le n_F$ and $n_{n,n^\prime}=0$ otherwise.
This allows the density matrix to be constructed from the 
first $n_F$ states incoming from the left and right to be written as
\begin{equation}
\label{1drho}
\rho_0(x,x^\prime) = \frac{1}{L} \sum_{n=1}^{n_F} \exp \left[ ik_n(x-x^\prime) \right]
                    +\frac{1}{L} \sum_{n=1}^{n_F} \exp \left[ -ik_n(x-x^\prime) \right].         
\end{equation}
Introducing the Wigner transformation term by term, the resulting 
Wigner distribution function is readily found to be 
\begin{equation}
\label{WFV0}
f_0(q,p) = \sum_{n=1}^{n_F} \left[ f_n(q,p_n) + f_n(q,-p_n) \right]
                                       = \frac{2\pi}{L} \sum_{n=1}^{n_F} \delta (p_n -\hbar|k_n|)
                                        +\frac{2\pi}{L} \sum_{n=1}^{n_F} \delta (p_n +\hbar|k_n|),
\end{equation} 
with $\delta(p)$ the Dirac delta function. 
Strictly speaking, as we consider wavefunctions that are
only non-zero on the scattering region, the delta functions should be
replaced by sinc functions of the form: 
$\frac{1}{\pi(k\pm p/\hbar)}\sin\left[ (k\pm p/\hbar)L/2 \right]$ that
approach delta functions for large $L$. In the case of large $L$, 
the model as described corresponds to fig.~\ref{fig1}a). 

Next, a potential step is introduced at $x=0$ to drive the system out of equilibrium and allow for a net 
current flow. The exact form of the scattering potential is not critical to the following
argument, but for ease of presentation we assume that the potential is varied over a small region $l << L$ 
allowing us to approximate the difference between the left and right electrodes as a potential energy step. 
For this case, the solutions to the one-electron Schr\"odinger equation may be written in scattering form
\begin{eqnarray}
\label{scattform}
\psi_{n}(x) =& \exp(ik_nx) + r\exp(-ik_nx) &  x < 0 \nonumber \\
\psi_{n}(x) =&  t \exp(ik_n^\prime x)      &  x > 0 . 
\end{eqnarray}
For spatially varying potentials centered at $x=0$ satisfying $l << L$, the asymptotic wavefunctions will 
likewise satisfy the scattering form and the following development remains valid with minor modification.
By implication, the energies for electrons entering the scattering region from the right are 
shifted by an amount given by the scattering potential height 
$\frac{(\hbar k)^2}{2m^*}\rightarrow \frac{(\hbar k)^2}{2m^*}+eV$. However the incoming momenta, as previously 
discussed, are unchanged. We apply a voltage greater than
the level spacings at the bottom of the reservoir conduction band where the energy density of states is greatest.  
With the introduction of the potential, the model for electrons entering from the left is the one-dimensional 
quantum mechanical scattering problem with a step-up potential. For electrons entering the scattering
region from the right, the problem is for an electron incident on a step-down potential. This is shown 
schematically in fig.~\ref{fig1}b) and the solution is well known for both cases and may be expressed in terms of the
transmission coefficients $T$. In this case, we can write the current for the system as
\begin{equation}
I\left[ V \ne 0 \right] = -\frac{e\hbar}{m^*L}\sum_{n=1}^{n_F} \left[ f_l(k_n) T_l(k_n;V) k_n - f_r(k_n) T_r(k_n;V) k_n \right] ,
\end{equation} 
with $T_l,T_r$ the transmission coefficients for electrons incoming from the left and right, respectively. Note
that $f_{l,r}(k_n\le k_F)=1$ and $f_{l,r}(k_n>k_F) =0$ for our example.
Time reversal symmetry requires that the left and right transmission functions for a given single 
particle energy $\epsilon$ are equal $T_l(\epsilon)=T_r(\epsilon)$, but the energies for the left and 
right states of equal momentum {\it are not} equal in the presence of a voltage.
This is seen by re-writing the transmission as functions of energy
\begin{equation}
I\left[V \ne 0\right]=-\frac{e\hbar}{m^*L}\sum_{n=1}^{n_F}\left[T_l \left( \frac{(\hbar k_n)^2}{2m^*};V\right) - 
                                                                T_r \left( \frac{(\hbar k_n)^2}{2m^*}+eV;V\right)\right] k_n,
\end{equation} 
resulting in a net current flow with application of voltage.

It can be shown that the Wigner transform of the reduced density matrix constructed from the scattering
wave functions satisfies the same open system boundary conditions as the zero voltage (plane-wave states) solution 
in the large $L$ limit and as we will demonstrate, approximately for finite values of $L$ typically used in
numerical studies. The electron reservoirs in this case are the regions outside of the central scattering site. The 
density matrix with $V\ne 0$ remains diagonal and we can again consider the Wigner transform term by term
\begin{equation}
\label{WTtermbyterm}
f_{n,n}(q,p) = \int_{-\infty}^{+\infty} \, dr\, \exp(-ipr/\hbar) \psi_{n}(q+\frac{r}{2})\psi^*_{n}(q-\frac{r}{2}).
\end{equation}
where now $\psi_n$ are wavefunctions of the scattering form eq.~\ref{scattform} on $[-L/2,L/2]$. 
We consider a point to the left of the scattering region $q_l=-L/4$ and find for
the Wigner distribution function 
\begin{equation}
\label{scattW}
f_{n,n}(q_l,p)=\frac{2\pi}{L}\delta(p-\hbar k_n)+\frac{2\pi}{L}r^2\delta(p+\hbar k_n)+\frac{4\pi}{L} r\,\cos(2k_nq)\delta(p)
\end{equation}
where in eq.~\ref{scattW} $r$ is the amplitude of the reflected component of the scattering wavefunction and the normalization
is fixed to that of an incoming plane wave. The Wigner phase space density at $q_l$ consists of the incoming 
momentum term at $p=+\hbar k_n$, the reflected momentum term at $p=-\hbar k_n$, and a zero mode term $p=0$ as 
discussed in ref.~\cite{CaZ83}. The Wigner transform is calculated at $\pm L/4$, in the middle of the electrodes,
to avoid coupling to the regions outside of $[-L/2,L/2]$ and to avoid interaction between the electrodes as voltage
is applied. If the point where the Wigner function is to be constrained is too close to the boundaries $\pm L/2$, 
the density matrix, as can be seen from the argument of the
Wigner transform eq.~\ref{WTtermbyterm} $\psi_{n}(\pm L/2+\frac{r}{2})\psi^*_{n}(\pm L/2-\frac{r}{2})$, will be zero. The vanishing of
the density matrix in this case is an artifact arising from truncating 
the wavefunctions outside of the scattering region. If the point where the Wigner function is to be constrained is calculated too close
to the scattering region, as voltage is applied the incident and reflected components of the scattering wavefunction mix with the
transmitted component, or in other words the two electrodes couple. To avoid coupling the electrodes, the Wigner function should be 
calculated within each electrode, but in a region avoiding interaction between the electrodes as a voltage is applied.

Again, as for eq.~\ref{WFV0}, for finite $L$ the delta functions in eq.~\ref{scattW}
should be replaced by sinc functions of the form $\frac{1}{\pi(k\pm p/\hbar)}\sin\left[ (k\pm p/\hbar)L/2 \right]$. 
For large $L$ when the sinc functions well approximate delta functions, eqn.~\ref{WFV0}
and \ref{scattW} satisfy the same condition for incoming momenta states. For finite $L$, the sinc functions centered
at $\pm k,0$ can overlap and the incoming momentum states can differ between the $V=0$ (plane-wave) and $V\ne0$ (scattering)
states. In fig.~\ref{fig3}a), we have calculated the Wigner function for a set of plane wave states incoming from the
left and the right with a Fermi momentum $k_F$ chosen to correspond to that of gold electrodes and a scattering region of
$L=2$ nm. A step potential of $V=0,1,2$ Volts is applied in the center of the scattering region and the resulting Wigner 
functions are displayed at $q_l=-L/4$ for each value of the applied voltage. These parameters have been chosen to compare
our analysis to typical calculations for molecular electronics, and in particular the calculation presented in ref.~\cite{DeG04a}. 
As voltage is applied, the incoming electron distributions are not exactly equal to
the $V=0$ distribution, but agreement is very close particularly for states near $k_F$ which provide the largest contributions to 
the current. The outflow of momentum is changed as a result of the scattering off the potential barrier and
this difference between the equilibrium ($V=0$) and non-equilibrium distributions ($V\ne0$) reflects the difference in
chemical potential between the electrodes. In fig.~\ref{fig3}b), the length of the region is taken to be $L=20$ nm and as seen
the incoming momentum distributions, even with wavefunctions only defined on the region $[-L/2,+L/2]$, remain essentially the 
same for $V=0,1,2$ volts. The momentum distribution drops sharply at $k_F$ well approximating the distribution shown in 
fig.~\ref{fig1}c). In fig.~\ref{fig3}c) and d), the Wigner distribution at $q_r=+L/4$ for $L=$ 2 and 20 nm, respectively, is given.
 In subsequent discussion, we assume a large value for $L$, but as fig.~\ref{fig3} indicates, the 
considerations apply well to electrode lengths as small as 1 nm. Also, the lowest $n_F$ states have been occupied 
corresponding to a $T=0$ distribution. However, there is nothing in the analysis that precludes the $V=0$ solution to be set to 
a thermal distribution and to constrain the solutions to the thermally occupied incoming states as a voltage is applied. If the 
scattering states with $V\ne 0$ are constrained to satisfy the Wigner function determined from the $V=0$ wavefunctions, 
the correct solution to the one electron Hamiltonian in the presence of a potential on the scattering region will be
obtained in the large $L$ limit and approximately for smaller values of $L$. Indeed it is observed that in the single particle 
case, constraining the incoming momentum inflow via the Wigner function implies solving the one-electron Schr\"odinger equation for a 
{\it specific} value of incoming momentum $p=\hbar k_n$, and otherwise results in the standard textbook presentation of one-dimensional 
quantum mechanical scattering.  


\subsection{A transport calculation with the single particle model} 

\begin{figure}
\includegraphics[width=5.2in,angle=0]{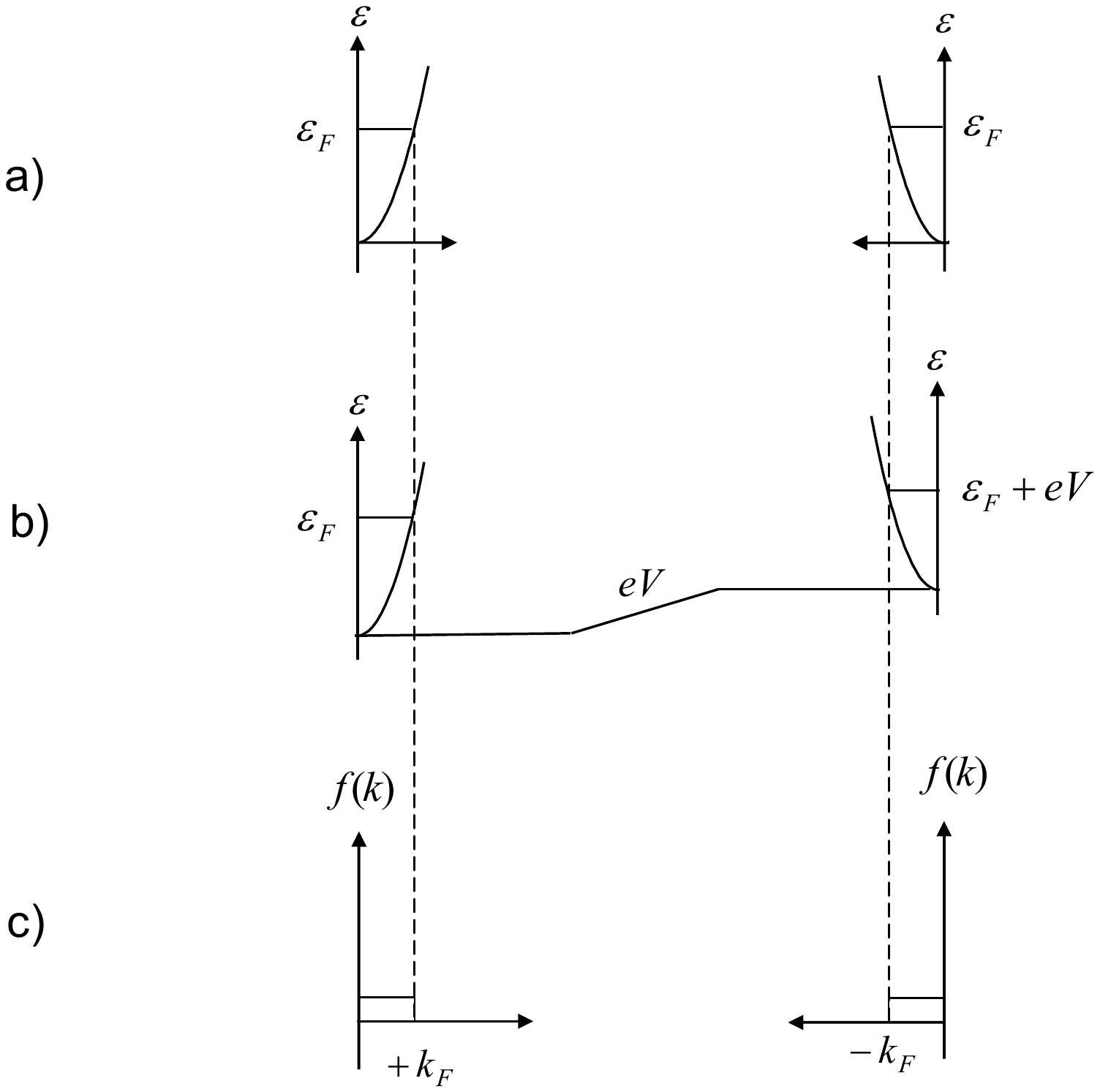}
\caption{ The Fermi-Dirac energy dispersions for incoming electrons
from two electrodes described by parabolic bands. a) with no voltage
difference between the electrodes, b) with applied voltage. c) The
corresponding momentum distribution functions for a) and b). Note that
there is no difference to the momentum distributions with application
of voltage.
}
\label{fig1}   
\end{figure}

\begin{figure}
\includegraphics[width=4.2in,angle=0]{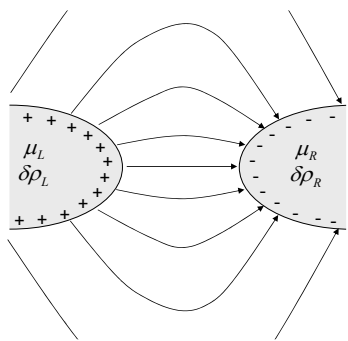}
\caption{A simple model of the action of two electrodes
in generating an electric field. The application of a
chemical potential difference results in surface charges
generating an electric field between the electrodes.
}
\label{fig2}   
\end{figure}

\begin{figure}
\includegraphics[width=4.2in,angle=0]{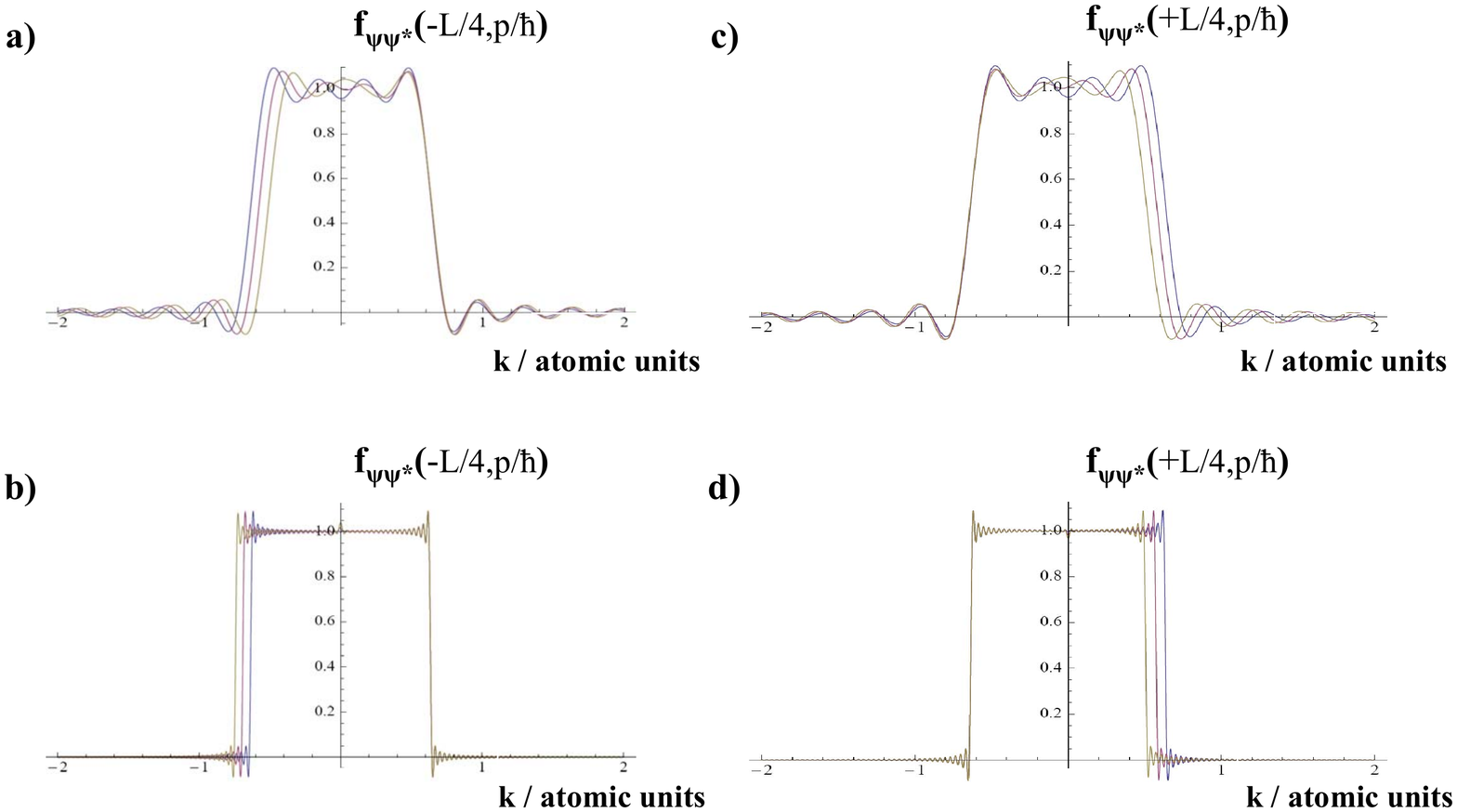}
\caption{The Wigner function calculated with scattering wavefunctions
defined on a region $[-L/2,+L/2]$ and with normalization chosen
such that a completely occupied state is with $f=1$. {\bf a)} The Wigner 
function calculated at $q_l=-L/4$ for a region of length $L=2$ nm
with $k_F$=0.12 nm$^{-1}$=0.635 a.u (corresponding to
gold electrodes) with $V$= 0 (blue), 1 (purple), 2 (green) 
volts. {\bf b)} Same as in a) but with $L$=20 nm. {\bf c)} The Wigner
function calculated at $q_r=+L/4$, otherwise the same as in
a). {\bf d)} Same as in c), but with $L$=20 nm. }
\label{fig3}   
\end{figure}

We again consider introduction of a small potential step to shift the energies of the states incoming from the right, as 
depicted in fig.~\ref{fig1}b). 
As voltage is applied, 
electrons incoming from the left with energies such that $\epsilon_{n_L}/e < V$ will see a potential step-up. The
number of these states is given approximately as
\begin{equation}
n_V \approx \sqrt{2em^*V}/\hbar\Delta k.
\end{equation}
For states incoming from the left, we approximate $T_l \sim 0$ for incoming energies
less than the potential step height, and $T_l \sim 1$ for energies greater than the potential
step height. In contrast, electrons incoming from the right see a step-down potential and 
we approximate $T_r \sim 1$ for all electrons incoming from the right. 
The electron current can then be estimated as
\begin{eqnarray}
I &\sim& -\frac{e\hbar}{m^*L} \left[ \sum_{n_l=n_V}^{n_F} k_{n_l} - 
                                   \sum_{n_r=1}^{n_F} k_{n_r} \right] \nonumber \\
  &\sim&  \frac{e\hbar}{m^*L} \sum_{n=1}^{n_V} k_n \nonumber \\
  &\sim&  \frac{e\hbar}{m^*L} \sum_{n=1}^{n_V} n \Delta k ,
\end{eqnarray}
with the convention that current flow is opposite the direction of electron flow. 
For small $\Delta k$ and large $n_F$ (these are standard conditions for derivation of the Landauer formula), we have
\begin{equation}
I\rightarrow \frac{e\hbar}{m^*L}\Delta k \int_0^{n_V} n\, dn = \frac{e^2}{h} V.
\end{equation}
The current and voltage yield a conductance $g_0=e^2/h$ and the Landauer result for conductance quantization
is obtained.

\begin{figure}
\includegraphics[width=4.2in,angle=0]{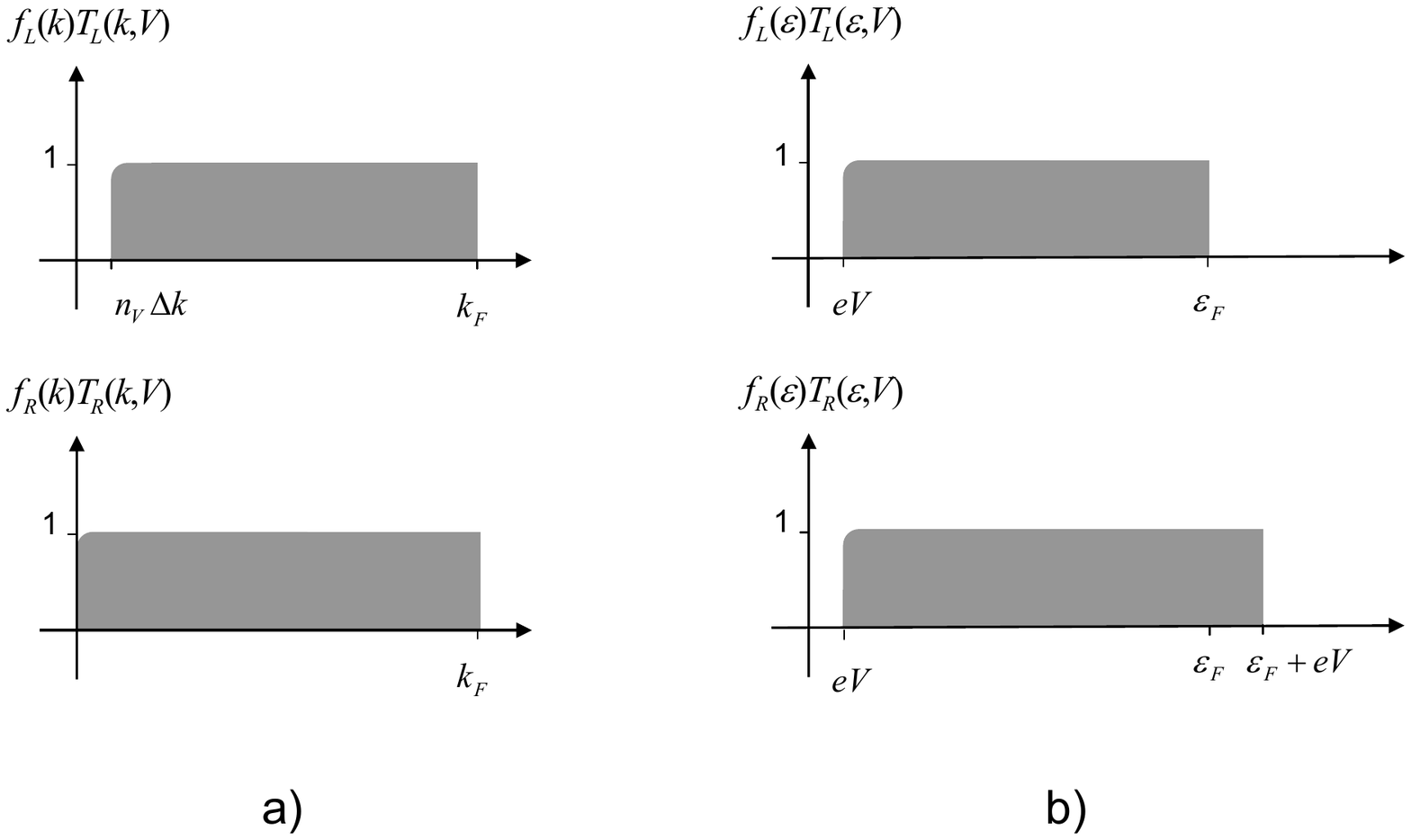}
\caption{a) The momentum distributions corresponding 
to the model of sect.~\ref{IV}. b) The energy distributions
for the model of sect.~\ref{IV}. Both distributions may be
used to calculated conductance quantization.
}
\label{fig4}   
\end{figure}

This derivation seems odd, as it appears that it is not the
states at the Fermi level that contribute to the current, but states low in energy (or momentum) that yield
a current. The situation can be summarized in fig.~\ref{fig4} . In fig.~\ref{fig4}a), the product of the momentum occupation
and the transmission coefficients for left and right states are given, whereas in fig.~\ref{fig4}b),
the corresponding product of the energy level occupations and transmission coefficients are given.
In fig.~\ref{fig4}a), it appears the currents arise from low momentum states, in fig.~\ref{fig4}b), currents appears to arise
from energy states at $\epsilon_F$. However, there is no contradiction. If the currents from 
momentum states with {\it energies} $\epsilon < \epsilon_F$ are summed, the currents associated with
states below $\epsilon_F$ cancel. 

The current incoming from the left can be rewritten as an integral over energy
\begin{equation}
I_l= -\frac{e^2}{h}\int_{eV}^{\epsilon_F} d\epsilon,
\end{equation}
using $\epsilon=\frac{(\hbar k)^2}{2m^*}$ and $d\epsilon=\frac{\hbar^2}{m^*}k\,dk$. Similarly the current from the
right is re-written but now with $\epsilon=\frac{(\hbar k)^2}{2m^*}+eV$
\begin{equation}
I_r= \frac{e^2}{h}\int_{eV}^{\epsilon_F+eV} d\epsilon.
\end{equation}
The resulting current is
\begin{equation}
I =  \frac{e^2}{h}\int_{\epsilon_F}^{\epsilon_F+eV}d\epsilon = \frac{e^2}{h} V,
\end{equation}
which will be recognized as the more familiar form for expressing conductance quantization. 
The physics is consistent whether calculating currents using the momentum 
distributions fig.~\ref{fig4}a) or the energy distributions fig.~\ref{fig4}b).

{\section{Discussion}\label{V}}

The first point of our presentation is that the MECS approach is consistent with a Landauer description of electron transport.
BK's conclusion ascribes the failure of their calculations for a
single particle and a correlated model to the MECS formulation of boundary conditions. The analysis of sect.~\ref{III}
highlights this is not the case; based on an analytical model, the use of the MECS formulation is consistent with a scattering 
approach to electron transport. The advantage of an analysis based upon an analytical model is that it avoids issues associated 
with linear response approximations, perturbation 
theory, variational methods in a finite basis, specific implementations of electronic structure, or other numerical approximations,
thereby allowing a clear focus on the physical assumptions made when using the method.  In sect.~\ref{IV}, we 
continue in this vein and demonstrate how the MECS formulation reproduces conductance quantization in a system of 
non-interacting electrons scattering off a potential barrier. 

We would also like to touch upon some formal points raised in the BK work. 
The authors criticize the use of a configuration expansion to describe transport problems. We note that
it is shown in several works that a properly designed variational calculation can provide accurate properties 
governing electron transport such as electronic spectra~\cite{GBG08} with compact expansion vectors~\cite{PMF01}. 
The variational structure of MECS calculations performed to date has also been noted by BK and we believe misinterpreted.
Using a variational approximation to the wavefunction does not result in an exact eigenfunction of the system Hamiltonian but 
rather the best approximation using the functional $<\Psi|H|\Psi>/<\Psi|\Psi>$ for the approximating function $|\Psi>$ and
subject to the application of the constraint conditions. 
As a consequence, it is well known that integrated quantities
such as the energy may be better approximated compared to local properties such as spin density or electron current density.
Hence in previous MECS calculations the possibility to introduce constraints to enforce current conservation on a scattering region 
was introduced. However, this is a numerical feature related to the variational nature of the calculations and the application of the open 
system boundary conditions does not imply the violation of current conservation contrary to a supposition in BK. We have already noted the 
origin of the current variations from variational calculations. 
It also well known that, for example, perturbation theories do not conserve many physically conserved quantities, including electronic current. 
Considerable care is needed in defining finite expansions that are current conserving~\cite{BaK61}. Hence we maintain the BK have incorrectly 
ascribed to MECS the curent oscillations they calculate in a tight binding model within linear response to the boundary conditions 
applied within MECS. 

As a final point, the application of the Wigner constraints in the case of a one-dimensional problem as we
have introduced and as attempted for a tight-binding linear chain in BK requires particular 
care in the following sense. Formally, for a single electron model of metallic electrodes (a reasonable assumption), 
the density matrix decays as $1/|x-x^\prime|^d$ with $d$ the spatial dimension of the electrode. 
How one decides to treat this long range behaviour influences the boundary conditions. Another way to express this is
that the density matrix does not satisfy Kohn's principle of `nearsightedness' for these examples~\cite{Koh96}, whereas in
three-dimensional models of metal electrodes~\cite{DeG04a,MGF09,FDG06,FaG07} the density matrix decays within typically less than
0.5 nm~\cite{ZhD01} thereby greatly simplifying the introduction of open system boundary conditions through use of the Wigner 
function. The decay of the density matrix in three-dimensional metallic electrodes helps in the calculation of the equilibrium 
($V$=0) Wigner function with small explicit electrode regions, without coupling between the electrodes or coupling the electrode 
equilibrium regions to the scattering region.
 
{\section{Conclusion}\label{VI}}

We have shown that the MECS boundary conditions and introduction of a chemical
potential imbalance between electrodes reduces to the correct single particle limit, and hence 
the claim in BK for the failure of the MECS boundary conditions to reproduce conductance quantization
is incorrect. The model analysis provided reveals that a failure of a calculation 
attempting to apply the boundary conditions using the Wigner function for momentum distributions 
for this or related models can not be attributed to a failure of the MECS formulation.

We have also provided an analysis on both the
formulation of many-electron scattering using the Wigner function boundary conditions,
and touched upon issues related to the numerical implementation 
of the model. We will present a similar analysis on more realistic models of atomic and
molecular scale systems and consider the effect of various numerical approximations 
on MECS transport calculations in future work.

{\bf Acknowledgments} This work was supported by a Science Foundation Ireland. We are
grateful to Stephen Fahy and Baruch Feldman for helpful comments on the manuscript.

\clearpage

\clearpage

\clearpage

\clearpage

\begin{thebibliography}{99}
\bibitem{SWF}
The use of SWF as terminology is misleading in the sense BK have introduced it: the 
time-independence of the Wigner distributions used to define the electrons within 
electrode regions has no more or no less significance than the time-independence of 
the Fermi-Dirac distributions used to constrain boundary conditions in the Landauer 
and related descriptions of electron transport.
\bibitem{note} If flux normalization and charge normalization in the case
of weak scattering are imposed, the degenerate solution of an incoming wave
from the right is suppressed without the imposition of the Wigner constraints
on the right. In realistic calculations for two electrodes, 
flux constraints on the right account for the momentum incoming from
the right.
\end{thebibliography}
\end{document}